\documentclass[letterpaper, 10 pt, conference]{ieeeconf}  % Comment this line out if you need a4paper

\IEEEoverridecommandlockouts                              % This 
\newcommand{\trtitle}{Facial Electromyography-based Adaptive Virtual Reality Gaming for Cognitive Training}

\usepackage{pdflscape}
\usepackage{tabularx,ragged2e,booktabs,caption}
\usepackage{subfigure}
\usepackage{multicol,tabularx,capt-of}
\usepackage{hhline}
\usepackage{multirow}
\usepackage{url}
\usepackage[most]{tcolorbox}
\usepackage{xcolor}

\usepackage{xcolor,colortbl}
\definecolor{Gray}{gray}{0.85}

\usepackage{graphicx}
\usepackage{amsmath}
\usepackage{amsfonts}
\usepackage{booktabs}
\usepackage{makecell}
\usepackage{balance}
\usepackage[nolist]{acronym}
\usepackage{amssymb}  % assumes amsmath package installed
\usepackage{cite}
\usepackage{bm}
\usepackage{moresize}
\title{\Huge \trtitle}

\author{Lorcan Reidy$^{1}$, Dennis Chan$^{2}$, Charles Nduka$^{3}$ and Hatice Gunes$^{1}$%\vspace{-5mm}% <-this % stops a space
\thanks{$^{1}$L. Reidy and H. Gunes are with the Department of Computer Science and Technology, University of Cambridge, U.K.
        {\tt\small lsr29@alumni.cam.ac.uk, hatice.gunes@cl.cam.ac.uk}}
\thanks{$^{2}$D. Chan is with Institute of Cognitive Neuroscience, University College London, U.K.
        {\tt\small dcfchan@doctors.org.uk}}%
        \thanks{$^{3}$C. Nduka is with Emteq Ltd, U.K. 
        {\tt\small charles@emteq.net}}%
}

\begin{document}
%\bstctlcite{IEEEexample:BSTcontrol}

\maketitle

\begin{abstract}
 %Dementia is a destructive phenomenon with an immense impact on society. There is an urgent need for research into new interventions for people suffering from dementia. 
 
 Cognitive training has shown promising results for delivering improvements in human cognition related to attention, problem solving, reading comprehension and information retrieval. However, two frequently cited problems in cognitive training literature are a lack of user engagement with the training programme, and a failure of developed skills to generalise to daily life.
 %
 %As life expectancy rises, age-related diseases causing dementia become more prevalent. In line with this, the health economic impact of dementia is escalating to unsustainable levels, with estimates that by 2050 dementia care will cost an annual \$1 trillion in the US alone. 
 %The development of interventions capable of improving cognition therefore represents an issue of the highest priority for healthcare. There has been considerable focus on cognitive training (CT) in particular, but work to date has been limited by two main factors, namely (i) the lack of transferability of CT gains to real life activities, and (ii) the lack of adherence to CT programmes. 
 %
 %
 This paper introduces a new cognitive training (CT) paradigm designed to address these two limitations by combining the benefits of gamification, virtual reality (VR), and affective adaptation in the development of an engaging, ecologically valid, CT task. Additionally, it incorporates facial electromyography (EMG) as a means of determining user affect while engaged in the CT task. This information is then utilised to dynamically adjust the game's difficulty in real-time as users play, with the aim of leading them into a state of flow. Affect recognition rates of 64.1\% and 76.2\%, for valence and arousal respectively, were achieved by classifying a DWT-Haar approximation of the input signal using kNN. The affect-aware VR cognitive training intervention was then evaluated with a control group of older adults. The results obtained substantiate the notion that adaptation techniques can lead to greater feelings of competence and a more appropriate challenge of the user's skills.
\end{abstract}


%\keywords{cognitive training; virtual reality; facial electromyography; affective computing; adaptive gaming.}


\maketitle
\section{Introduction}
%Dementia occurs as an consequence of severe brain damage, often due to neurodegenerative disorders such as Alzheimer's disease, and is defined as a deterioration in memory and thinking to the point where activities of daily living are impaired and functional independence is lost. As dementia prevalence rises with the ageing population, attention is turning increasingly towards interventions that may ameliorate cognitive decline in patients in earlier stages of disease, prior to the onset of dementia.

Cognitive training (CT) has garnered considerable attention due to the promising intervention outcomes it has provided for improving cognitive skills such as attention, problem solving, reading comprehension and information retrieval. Clinician-delivered or computerised CT has been used for children with attention problems (ADHD)
\cite{MooreEtAl2018} and engagement in intellectually stimulating lifestyle activities has been observed to help maintain cognitive function into later life, via enhancement of cognitive reserve \cite{Stern2012}. The aim of CT therefore is to deliver such benefits to cognition as a targeted discrete intervention. However, to date, the numerous studies and CT products, often commercially marketed as "brain training" programmes or apps, have not been found to produce clear evidence of benefit to cognition  \cite{GatesEtAl2019}. 
This failure has been attributed to several factors. The first problem (P1) 
 is that of adherence to the CT programme; with traditional CT programmes often having attrition rates exceeding 15\% \cite{WykesEtAl2011}. Prior studies have indicated that participant adherence can decrease with increasing intervention complexity and intensity \cite{LamEtAl2015}, \cite{ColeyEtAl2019}, with inhibition due to poor task performance being a strong predictor for training dropout \cite{Arbiv2015}. These CT programmes are overly repetitive and fail to address the motivational deficits often characteristic of older people with memory difficulties \cite{SavulichEtAl2017b}. Furthermore, it has been suggested that a VR format may increase training adherence (when compared to an on-paper equivalent task) in individuals with Mild Cognitive Impairment (MCI) and dementia patients \cite{ManeraEtAl20106}.
%is that of adherence to the CT programme; participant engagement often drops off after a period because they become bored or frustrated with the training programme. This negative effect, and the resultant lack of engagement, is usually a result of the training not being sufficiently challenging, being too difficult, or simply not being sufficiently compelling to begin with 
%\cite{Simons2016}. 
%
The second problem (P2) is that the skills developed and the cognitive improvements made during these training programmes do not transfer to daily life. While the user might get better at playing the games, there is little evidence that these skills will generalise beyond that \cite{Simons2016}. Existing dynamic difficulty adjustment (DDA) models have historically relied on the player’s performance to infer what changes in difficulty are appropriate. However, Pagulayan et al. point out that a game’s evaluation factor ought to be the affective experience provided by the gameplay, rather than user performance as in productivity software \cite{PagulayanEtAl2002}. Mishra et al. were also of the opinion that the input channel for closed-loop games should include real-time data from player interactions and behaviour, rather than just player performance metrics \cite{MishraEtAl2016}.
%The third main problem relates to the lack in many studies of an appropriate "active control" against which the CT intervention can be compared, thus compromising interpretation of study results.

The work presented in this paper introduces a novel CT programme designed to address these two problems, with an ultimate goal of employing this CT environment for people with early stages of dementia to support cognitive training. In keeping with other approaches to CT, a game-based paradigm is used to enhance participant enjoyment and thus increase adherence (addressing P1). The use of VR to create simulated real-world environments within which the CT game is enacted helps overcome the transfer problem and facilitate extension of CT-generated gains to real life activities (addressing P2). Additionally, the design of the CT task incorporates an active control arm and  an affective feedback loop that enables the dynamic adjustment of the game's difficulty (DDA), in real-time, based on the player's affective state \cite{Liu2009} and their in-game performance (addressing P1). The affective feedback loop is based on measuring facial electromyography (EMG) signals related to the nonverbal facial behaviour of the participants, and training supervised machine learning models with user self-reported affect labels. The employment of facial EMG is due to three reasons: (i) other physiological measures such as the galvanic skin response (GSR) have been reported to provide extremely noisy signals in gaming where the user is using a controller and moving around  during the game play \cite{GabanaEtAl2016}; (ii) it is widely acknowledged that the face is the primary means for communicating affective and cognitive mental states, including emotions, interest, agreement, comprehension, concentration and intentions \cite{Baron-Cohen96}, \cite{CohnEkman2005}; and (iii) computer-vision based analysis of facial expressions, despite its efficacy \cite{Sariyanidi2015Automatic}, \cite{Martinez2019Automatic}, is not appropriate in a VR context as the user's face is mostly occluded by the head-mounted display.

\section{Related Work}

\subsection{Cognition, Memory and Affect}

%Memory can largely be understood as information gained from past experience, that can be used in the service of current, or future, adaptive behaviour. Prominent literature in the field has led away from the unsubstantiated perception that memory functions as a unitary process towards the concept that memory can be categorised into subsystems (operating across different regions of the brain), and that its functions are underpinned by the interactions between these subsystems \cite{schacter1994a}. These subsystems are classified according to their content and function. 

The bisection of memory into short-term-memory (STM) and long-term-memory (LTM), first proposed by Hebb \cite{hebb1949a}, represents the core underlying division imposed in the taxonomy of cognitive systems. 
%
%LTM has an indefinite capacity and retention period, and is unbounded by context and task demands. STM has a limited capacity (typically around 7 items can be stored at a time 
%\cite{miller1956a}) and duration (storage is fragile, information can easily be lost with distraction or passage of time), and stores information in a task dependent manner (e.g. in terms of physical qualities of the experience - such as what we see, do, hear etc.).
%These memory systems can be further divided into a more granular taxonomy. The unitary notion of STM can be partitioned to recognise a distinct cognitive system, working memory (WM), that is responsible for temporarily storing task relevant information available for manipulation. 
%WM is a variant of the unitary STM concept, and is comprised of three subcomponents: (i) the central executive, which controls attention and is particularly susceptible to the effects of Alzheimer's disease 
%\cite{baddeleysquire1992a}; (ii) the visuospatial sketch pad, which stores, and manipulates, visual and spatial information; and (iii) the phonological loop, which maintains auditory information through rehearsal. 
%
%
%These memory systems can be further divided into a more granular taxonomy. 
The unitary notion of STM can be partitioned to recognise a distinct cognitive system, working memory (WM), that is responsible for temporarily storing task relevant information available for manipulation.
LTM can be divided into procedural (non-conscious) memory, which underpins skills/habits/conditioning, and declarative (consciously accessed) memory for storing facts and events 
\cite{squire2004a}. Declarative memory can be further separated into semantic memory (SM), a store of facts about the world, and episodic memory (EM), the facility for re-experiencing events in the context in which they originally occurred 
\cite{tulving1983a}. EM is widely considered as a cognitive competence unique to humans. Unlike SM, it explicitly encodes spatial, contextual and temporal information. 
%It therefore requires the engagement of areas of the brain in addition to those that support SM.
%

Neurological disorders such as attention-deficit/hyperactivity disorder (ADHD) or dementia are associated with memory impairments. Numerous studies have shown deficits in short-term memory, long-term memory and coordination of multiple tasks resulting from such disorders (e.g. see \cite{baddeley1991a}). Therefore, the VR cognitive training intervention developed for \textit{this work incorporates both WM and EM training tasks}. 

WM capacity can be tested through a variety of tasks. These tasks typically come in the form of a dual-task paradigm that combines a measure of memory span (a STM test that involves immediate recollection of an ordered list of items) with a simultaneous processing task (e.g. see \cite{daneman1980a}). It has also been argued that WM reflects the ability to maintain multiple, task-relevant, pieces of information in the face of distracting irrelevant information \cite{engle1999a}. The WM training task implemented in this work draws on both of these ideas. 
%Due to the slightly broader definition for EM, and the numerous competing theoretical and empirical perspectives on it, 
A wide variety of methods have been developed to assess EM capacity, but not all produce consistent results \cite{cheke2013a}. For the purpose of this work, the implemented EM training primarily focuses on spatial memory (SM) tasks, as they engage the same regions of the brain that EM requires \cite{burgess2002a}. SM tasks also proved to be a good fit for the VR paradigm, where ecologically valid scenarios could be presented to the user in an engaging manner.

Bennion et al. \cite{bennion2013a}, in their study of the effect of emotion on memory, suggest that there is strong evidence to support the following hypotheses: emotion usually enhances memory; when it does not, its effect can be understood by the magnitude of elicited arousal (with arousal benefiting memory up to a point, but then having a detrimental influence); and when emotion facilitates the processing of information, it also facilitates the retention of that information. 
%However, they also outline, and emphasise the importance of, limiting factors on the effects of emotion on memory. They reference the results of Waring and Kensinger's study, which found that patients with Alzheimer's disease tended to show little-to-no enhancement of emotional memory within a laboratory setting, but they were more likely to remember emotive events in real-life [66]. They qualified the second hypothesis by saying that, while the hypothesis appeared to hold, the effects of arousal quality (e.g. excitation vs. agitation) were not sufficiently understood yet.
The general notion that arousal will enhance memory, up to a point, was reinforced by Yeh et al. in their research of the effects of negative affect on WM capacity 
\cite{yeh2015a}. They promote the idea of a game that appropriately challenges users in order to activate their attention, while avoiding negative emotional responses. 
%This phenomenon has also been observed in a physiological context. Suriya-Prakash et al. found, in their study on the influence of visuospatial WM tasks on heart rate variability (HRV), that HRV was lower among poor WM performers compared to good performers \cite{suriya-prakash2015a}.

\subsection{Affect and Gaming}
%Affect recognition is a complex problem, with expressions for the same emotions varying across individuals (depending on culture/personality/disabilities etc.). 
%A pivotal theory, which the development of affect aware systems follows from, is that the perceived physiological changes that occur can be interpreted as the emotion itself rather than just its expression [34]. Affective computing systems can therefore identify emotions by analysing the pattern of physiological changes in a user. While a significant amount of information can be derived from physiological signals , the intrusiveness of applying physiological sensors has been detrimental to the viability of such techniques in real-world scenarios. This has been mitigated to some extent with recent developments in the design of wearable sensing devices (for examples of sensors used in a real-world context see [35]). A promising avenue for the development and adoption of affective systems, which is touched on in this research, is the (seamless) integration of physiological sensors with equipment that already sees widespread use (e.g. VR headsets, see Fig. 1).
Two domains, particularly relevant to this work, have shown promise in recent literature for the application of affective computing methods. The first is the domain of cognitive training, which is motivated by the strong relationship between emotions and cognitive performance 
\cite{gabana2017a}. The second is the domain of video games, where the interaction has been noted as a predominantly emotional one \cite{yee2007a} and therefore susceptible to affective adaptation -- i.e., dynamically changing the gameplay experience based on affective signals read from the player. 

%Video games, perhaps more than any other entertainment medium, engage users in a diverse range of experiences. Players' motivations for engaging with video games vary. They have been categorised into three overarching motives \cite{yee2007a}: players seeking game mastery and competition (achievement), players who want to interact with others and develop in-game relationships (social), and players who pursue escapism by engaging with a game's story (immersion). These motivations indicate that the activity of playing video games is a predominantly emotional one. This suggests that game development is a domain that would benefit from the application of affective computing techniques.
%The research and development of affective video games represents an effort to increase the emotional bandwidth of games (e.g. consider a scenario where the game affects the player, and the player also affects the game) [8]. 
Affective gaming can be realised through the use of biofeedback techniques. However, for a game to be considered affective (and not simply a biofeedback game), it must exploit this biological information to propagate affective feedback 
\cite{bersak2001a}. That is, the game is an intelligent participant in the biofeedback loop. What distinguishes affective feedback from biofeedback is that the player is not deliberately controlling their physiological responses in order to influence gameplay.

There are numerous novel possibilities for emotive twists on conventional gameplay experiences. For example, Reynolds and Picard developed AffQuake \cite{reynolds2004a}, a modification to ID Software's Quake II that incorporated affective signals to alter gameplay in a variety of ways (e.g. in StartleQuake, when a player becomes startled, their avatar also becomes startled and jumps back). Valve Corporation have also experimented with similar modifications to their games: Half-Life 2 and Left 4 Dead 2 
\cite{bouchard2012a}. In these modifications, the player's stress level, measured as the electrical response of their skin, determines the pace of the gameplay.
Compared to conventional gameplay, the use of VR in video games induces a greater degree of engagement and immersion in players. This heightened immersion, named presence (or the feeling of being there in the virtual world), has been reported to directly impact the affective states experienced by the player during gameplay -- i.e., high levels of presence induce more intense and vivid emotions \cite{riva2007a}. 

\subsection{Therapeutic Applications of Virtual Reality}
%
%This perceived link with affective states has led to VR being referred to as an 'affective medium', or as an 'empathy machine' (technology with the capability to make sensible to oneself the emotional experience of another \cite{bollmer2017a}). 
%The emotive nature of VR experiences naturally leads to a synergistic relationship with affective computing techniques \cite{shumailov2017a}.
%
%VR environments are being utilised more and more in neuroscience research and in behaviour therapy interventions. 
%The key advantage offered by VR in the neuroscience field is the ability to place patients in ecologically valid, safe and controlled environments that provide multisensory stimulation \cite{bohil2011a}. Successful therapeutic applications of VR have ranged from desensitisation treatments for PTSD sufferers to serving as a tool for pain distraction in forms of exposure therapy for people suffering from various phobias and anxiety disorders \cite{denmark2017a, pedroli2018a, gabana2017a, sanchez-vives2005a}. %
%
%It has also seen recent use in novel cognitive tests and training interventions 
%
%
%The level of control afforded by virtual reality cognitive training (VRCT) enables the gamification of cognitive tasks. Its goal is to motivate users to engage more with, and perform better at, whatever task it is being applied to. 
%
%
%A subset of literature in this domain, that has recently seen an upsurge in attention, has been on the use of VR in creating ecologically valid experiments. 
In a systematic review of computerised cognitive training literature, Hill et al. concluded that computerised CT is a viable intervention for enhancing cognition in people with MCI \cite{hill2017a}. Interestingly, they found that for individuals with dementia, the only clinically meaningful effect sizes were found in studies that utilised immersive technology such as VR or the Nintendo Wii %(which has motion controls similar to VR interaction methods) 
\cite{hill2017a}. 
%Teo et al. in their literature review of VR as a platform for neurorehabilitation, reported that established evidence supports the efficacy of VR, but suggested that the combination of VR and conventional therapies is likely to be more efficacious compared to using either alone \cite{teo2016a}.
%
%
%For the purpose of this work, gamification is understood to include the use of gameplay mechanics and interactions. 
%Recent literature has reported on successful applications of gamification with older adults, with focus groups made up of older adults expressing an acute awareness of the need to strengthen their cognitive skills and regarding games as a means to do so \cite{kayali-a}. 
%A number of design considerations have been outlined that should be taken into account when targeting this population \cite{kayali-a}. These considerations emphasise the accessibility of the game (e.g. limiting game speed, use of strong contrasting colours etc.), suggest the use of adaptive difficulty schemes, and encourage careful selection of physical interaction methods \cite{miesenberger2008a}. 
%Additionally, Whitlock et al. recommend the provision of training support prior to getting started, and during early gameplay sessions \cite{whitlock2010a}.
%
These findings strongly support the idea that greater transference of cognitive performance to daily life could be achieved through training on ecologically valid tasks in VR. 

%This would be in contrast to the most prevalent ‘brain-training’ products on the market. Simons et al. strongly contest that there is adequate evidence to suggest that these popular products  (specifically Nintendo’s “Brain Age” and its sequels \cite{Nintendo:2012}, Lumos Labs’ “Lumosity” \cite{Lumosity:2020}, Posit Science’s “BrainHQ” \cite{BrainTraining:2020}, and Cogmed \cite{Cogmed:2020} train cognitive skills that will generalise to real-world cognition. They highlight that supporting studies for these products have tended to only show narrow transfer (the benefits, when present, have applied almost exclusively to other laboratory tasks) \cite{yee2007a}.
%
%A number of novel CCT applications have been developed with the aim of improving on some of the cited shortcomings in existing CT programmes.
%
Savulich et al. attempted to address the motivational deficits in older populations with memory impairments through a novel memory game on an iPad “Game Show” which targeted EM as its cognitive process \cite{SavulichEtAl2017b}. The game was designed to have the same motivational properties typical of computer games (tutorials, stimulating music, progression, etc.) and employed DDA with in-game performance as a sole input. They conducted a randomised controlled trial with 42 patients (aged 45 and over) diagnosed with amnestic mild cognitive impairment, splitting the participants evenly into a training and control group. 
Their results showed significant EM improvements in the CT group, and suggested that gamification enhanced participant motivation.

Gamito et al. investigated the effectiveness of VR for neuropsychological rehabilitation \cite{GamitoEtAl2015} by developing a VR application for CT which incorporated attention and memory tasks resembling daily real-world activities. These tasks targeted WM, visuo-spatial orientation, selective attention, recognition memory and calculation. The training linearly increased in difficulty throughout the sessions. This was employed in a study consisting of twenty stroke patients, with a mean age of 55 years (SD = 13.5), who were randomly assigned between an intervention and control group. The results of their study generally supported the efficacy of VR-based interventions for CT, with significant benefits to memory and attention functions observed in the intervention group (with the exception that no significant improvement was seen in visual memory).

Optale et al. conducted a 6-month long randomised controlled pilot study investigating the efficacy of their VR training intervention in lessening cognitive decline and improving memory functions \cite{OptaleEtAl2010}. The VR tasks asked the user to navigate a simple environment and tested their capacity to recall the route they have taken and their orientation, with a gradual, linear, increase in the complexity of the stimuli. They recruited 36 elderly individuals (median age 80 years) with memory impairment and divided them into an experimental group and an active control group (which received music therapy sessions). Their results showed that participants in the experimental group exhibited improvement in general cognitive functioning and verbal memory, with the most significant effects observed in long-term memory. 
%Improvements in executive functioning were much less significant, and no effect was observed for visuospatial abilities. In contrast to the experimental group, participants in the active control saw progressive cognitive decline.

Despite their promising results, none of the reviewed studies utilised affective feedback as part of their VR intervention, which is the main focus of this work.  
There also exist other VR applications for CT (e.g., \cite{pedroli2018a}, \cite{DonigerEtAl20108}, \cite{CaggianeseEtAl2018}), however, these have not yet reached the point of carrying out studies with their target population.

\subsection{Emotion Sensing from Facial EMG}
%facial exp
%The face is widely considered to be the most reliable source of affective information. It conveys information about a person's age, sex, background and identity, what they are feeling, and what they are thinking \cite{ekman2005a}. 
%There are two prominent strategies for measuring facial expressions. The first, and most prevalent, method relies on computer vision and image processing techniques 
%\cite{bartlett1999a, sebe2007a, koelstra2010a, littlewort2016a}. 
%It typically involves three steps: face detection in an image or video stream frame, facial feature extraction, and facial expression (or Action Unit \cite{ekman1978a}) classification. 
Computer-vision based analysis of facial nonverbal behaviour (facial actions or facial expressions) is not appropriate in a VR context as the user's face will be mostly occluded by the  head-mounted display (HMD). 
Cohn and Ekman in \cite{CohnEkman2005} provide a detailed review about early studies that have used surface EMG to measure facial muscle activity in relation to emotion and the evidence found in terms of predictive correlation with self- and observer reported emotion. 
Therefore, this work utilises facial electromyogram (EMG), recorded from surface electrodes placed over regions of the user's face, to measure signals related to the affective facial behaviour of the participants, using a novel device, Faceteq \cite{mavridou2017a}.
%Measuring Facial EMG benefits from high sensitivity, enabling the detection of slight muscular movements that may not be evident to the human eye. However, a significant drawback is that the application of facial surface electrodes is obtrusive to the user, and makes the user aware of, and self-conscious about, their facial expressions and the measurement thereof \cite{ekman1992a}. This work aims to negate this drawback through the use of a novel affective human-computer interaction device, Faceteq \cite{mavridou2017a}. 
%This is a prototype device that enables the sensing of upper facial expressions through electromyograms, while wearing a VR HMD.
%Most of this literature has used facial EMG to discriminate between positive and negative emotion (Cacioppo, Petty, Losch, & Kim, 1986).

%
For extracting features from facial EMG, Jerritta et al. investigated the application of higher order statistics (HOS) to derive a set of facial EMG features for classifying Ekman's six basic emotional states \cite{ekman1992b}. 
They used audio-visual (video clips) stimuli to induce emotional responses in participants, and employed a kNN classifier with PCA as a dimensionality reduction technique. 
%They found that the HOS features outperformed commonly used statistical features (mean absolute value (MAV), standard deviation, etc.)\cite{hamedi2012a}.
%, though it is worth noting that this did not include commonly reported informative temporal features such as root mean square (RMS) and integrated EMG (IEMG) \cite{hamedi2012a}. 
Their results showed that the use of PCA prior to classification improved the classifier's accuracy, achieving an average classification rate of 69.5\% across the six basic emotions (anger, disgust, fear, happiness, sadness and surprise).
%(using a 70/30 training/validation split of the data for CV).
%
Perusquia-Hernandez et al. \cite{perusqu2017a} investigated the recognition of spontaneous vs. posed smiles, using spatial and temporal patterns of facial EMG. 
%Spontaneous smiles were elicited through audio-visual stimuli, while posed smiles were requested by the experimenter. 
%(with participants informed of the purpose, that being to classify EMG signals). 
Due to the unbalanced nature of the collected data, they undersampled the majority class to match the minority class samples (as in \cite{shumailov2017a}). The best classification results for this 2-class problem, obtained using spatial-temporal features with a Gaussian kernel SVM, range from 85.23\% to 96.43\% (across participants), using a 70/30 training/validation data split.
Soon et al. developed an application for speech recognition based on facial EMG. Three participants were asked to say a series of numeric (spoken in Malay and English) and command words (spoken in English). Temporal features were extracted from a DWT-Haar approximation of the input signal. Four different classifiers were evaluated: Random Forest, Linear Discriminant Analysis, Naive Bayes and Multilayer Perceptron. Classification results were obtained through a cross-validation scheme with a 66/34 training/validation data split. Random Forest provided the best overall performance with temporal features achieving 64.7\%, 49\%, and 41.8\% in Malay, English, and command words respectively.

\section{Study Design}
\subsection{Game Design}
%The game was developed for the HTC Vive VR platform, using C# and the Unity  3D game engine. 
%The HTC Vive (pictured in Fig. 12 with attached Faceteq sensor) is designed to track the player, enabling free movement through physical space. The player interacts with the game environment through two motion sensing controllers. These interactions include picking up game objects, selecting game objects with a laser pointer and, to facilitate navigation across greater distances in the environment, laser-guided teleportation. 
Two separate virtual environments were developed, a virtual supermarket and a virtual multi-room museum (see Fig.~\ref{fig:supermarket} and Fig.~\ref{fig:museum}). These locales provided the setting for the WM and EM tasks respectively, and were selected to promote the ecological validity of the intervention, i.e., both environments are likely to be familiar to the older target population and, in the case of the supermarket, to reflect a daily activity. The underlying hypothesis was that 
%if existing, more abstract, cognitive training interventions consistently improve performance on the training task, and closely related tasks, but fail to improve real-world performance [6], then 
by setting the tasks in highly immersive virtual re-creations of real-world environments and having users perform practical tasks (e.g. collecting products from a shopping list and interacting with displays in a museum) the acquired cognitive skills would better generalise to daily life.

In the initial phase of the work, a fixed difficulty framework was implemented for both the WM and EM tasks consisting of three difficulty levels (easy, medium and hard). 
%This framework consisted of three difficulty levels (easy, medium and hard), 
This was designed with the purpose of eliciting a range of emotional responses from study participants and, thus, generating a balanced dataset. These difficulty levels differed in the cognitive load required from the participant (e.g. shorter/longer shopping lists in the supermarket, less/more display locations to remember in the museum).
%%%%%%%%%%%%%%%%%%%%%%%%%
%\begin{figure}[!t]
%\begin{tabular}{c}
%	\includegraphics[scale=0.12]{AffectiveSlider.png}\\
%	\end{tabular}
%    \caption{The Affective Slider, Betella and Verschure's digital self-assessment scale for the measurement of human emotions \cite{betella2016a}. The top slider indicates level of arousal, the bottom slider indicates valence of emotion.}
%	\label{fig:slider}
%	 \vspace{-4mm}
%\end{figure}
%%%%%%%%%%%%%%%%%%%%%%%%%%%%%%%%%%%%%%%%%
%
%Fig. 6.	The Affective Slider, Betella and Verschure's digital self-assessment scale for the measurement of human emotions \cite{betella2016a}. The top slider indicates level of arousal, the bottom slider indicates valence of emotion.

Both tasks employ the same annotation and EMG logging scheme. EMG data is recorded from when the task starts. After every 45 seconds of gameplay (a time period arrived at through pilot tests) the recording is paused and written out to a log file (created for that segment). When this occurs, the game environment fades to black and the affective slider 	(see Fig.~\label{fig:slider}) (implemented in VR to mitigate gameplay disruption), is displayed to the player. Players interact with the slider using a VR laser pointer and, when they are happy with their selection (which should best describe the average affect experienced by the player during the preceding 45 seconds of gameplay), press a confirmation button to append the arousal/valence values to the associated EMG log. Gameplay and EMG recording then resumes. This process repeats until the timer runs out. 

Putze et al. performed a systematic investigation into the effects of interrupting the VR experience through a questionnaire either inside or outside of VR \cite{Putze-2020}. Their results showed that administering questionnaires in VR reduces the Break in Presence without affecting the self-reported player experience. This motivated our decision to collect the labels during gameplay. The use of the affective slider was motivated by its facility for expeditious annotation and its interpretability. We therefore decided that obtaining these labels inside VR will better represent the range of emotions experienced at different points during gameplay (while the experience is still fresh).
%with the added benefit of the elimination of potentially time consuming post-game annotation sessions. 
Score tracking and a leaderboard (staples of gamification) were also included for both tasks.
%
%
%in-experience surveying in VR may break the sense of presence  \cite{UsohEtAl-2000}
%These mechanics have been shown to improve engagement \cite{hamari2014a} and serve as progress indicators, guiding and enhancing player performance \cite{mekler2013a}.
% The leaderboard is displayed to the player at the start screen (prior to task commencement) where they can see the scores achieved by previous players, giving them something to aim for during their session. The (updated) leaderboard is then displayed at the end of each round, allowing the player to see how they performed relative to others. This was done to evoke more intense emotional responses during gameplay (e.g. satisfaction/frustration from positive/negative performance), thereby obtaining a more balanced dataset. 

A heads-up-display (HUD) was included that enabled the player to track their current score, time left and other task specific information. Together, the inclusion of these elements was intended to draw the player's focus away from the novelty of VR (thereby mitigating the expected positive bias in the dataset) onto their task performance. To further associate player performance and emotional response, audio-visual stimuli were added in response to correct (bell ringing sound and confetti explosion) and incorrect (buzzing sound and red X) answers.

\subsubsection{Working Memory Task}
%%%%%%%%%%%%%%%%%%%%%%%%%
\begin{figure}[!t]
	\begin{tabular}{c}
		\includegraphics[scale=0.18]{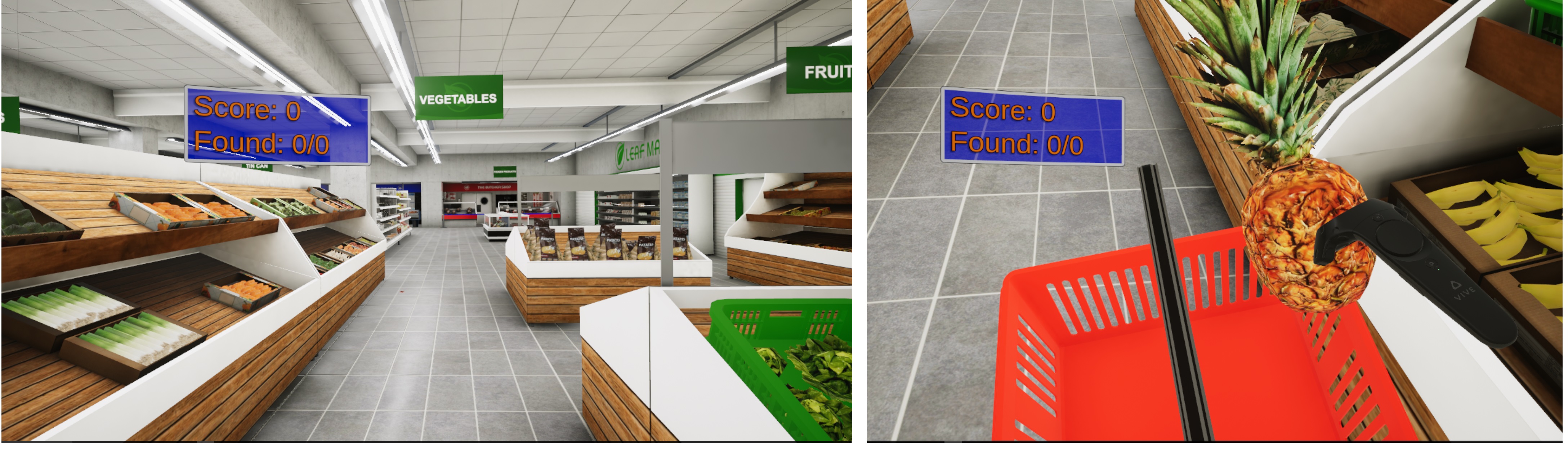}\\
	\end{tabular}
	\caption{The custom virtual supermarket environment, in which WM tasks were carried out, and player interaction with this environment. }
	\label{fig:supermarket}
	 \vspace{-4mm}
\end{figure}
%%%%%%%%%%%%%%%%%%%%%%%%%%%%%%%%%%%%%%%%%
The goal of the WM task is to find a (randomly generated) array of products in a virtual supermarket (see Fig. \ref{fig:supermarket}), placing each product in a shopping basket. The products are specified to the player at the start of each round through the HUD. Each product on the shopping list is displayed on the HUD (as an image and text description) for 1 second, with a 500 ms interval. Players are challenged to remember remaining items on the shopping list (stored in WM), while they actively search for each product. 
%If players find all the items on the list with time to spare, a new list is generated and they can continue collecting items to increase their score (so there is some incentive to perform the tasks quickly).
The number of products to be collected is determined by the difficulty level. The medium difficulty tasks users with finding 7 products, this is intended to be the most engaging and balanced difficulty for most users (based on Miller's magic number seven, plus or minus two \cite{miller1956a}). The easy and hard difficulties task users with finding 2 and 12 products respectively. These difficulties were designed to increase the likelihood of inducing negative affect in users, i.e., calm-negative on easy (bored due to insufficient challenge) and energetic-negative on hard (frustrated due to excessive challenge). The specific number of products for each difficulty level was determined through pilot testing and feedback.
%Fig. 8.	Player interaction with the environment. 
%The shopping basket is held in one hand while the player reaches for various shopping products/groceries and places them in the basket.
%Players hold a shopping basket in one hand (tied to the location of one controller), and interact with the supermarket environment by reaching out and grabbing items with their other controller and placing them in the basket (Fig. 8). 
For each correct item collected, 5 points are added to the player's score. Collecting an item that was not on the shopping list reduces the player's score by 4. Therefore, while the player is incentivised to carry out the task quickly, the priority is to ensure that no mistakes are made. The random generation of shopping lists promotes the task's replayability, maintaining the emphasis on short-term WM (rather than remembering the shopping lists from previous attempts) on repeated playthroughs.

\subsubsection{Episodic Memory Task}
%%%%%%%%%%%%%%%%%%%%%%%%%
\begin{figure}[!t]
	\begin{tabular}{c}
		\includegraphics[scale=0.15]{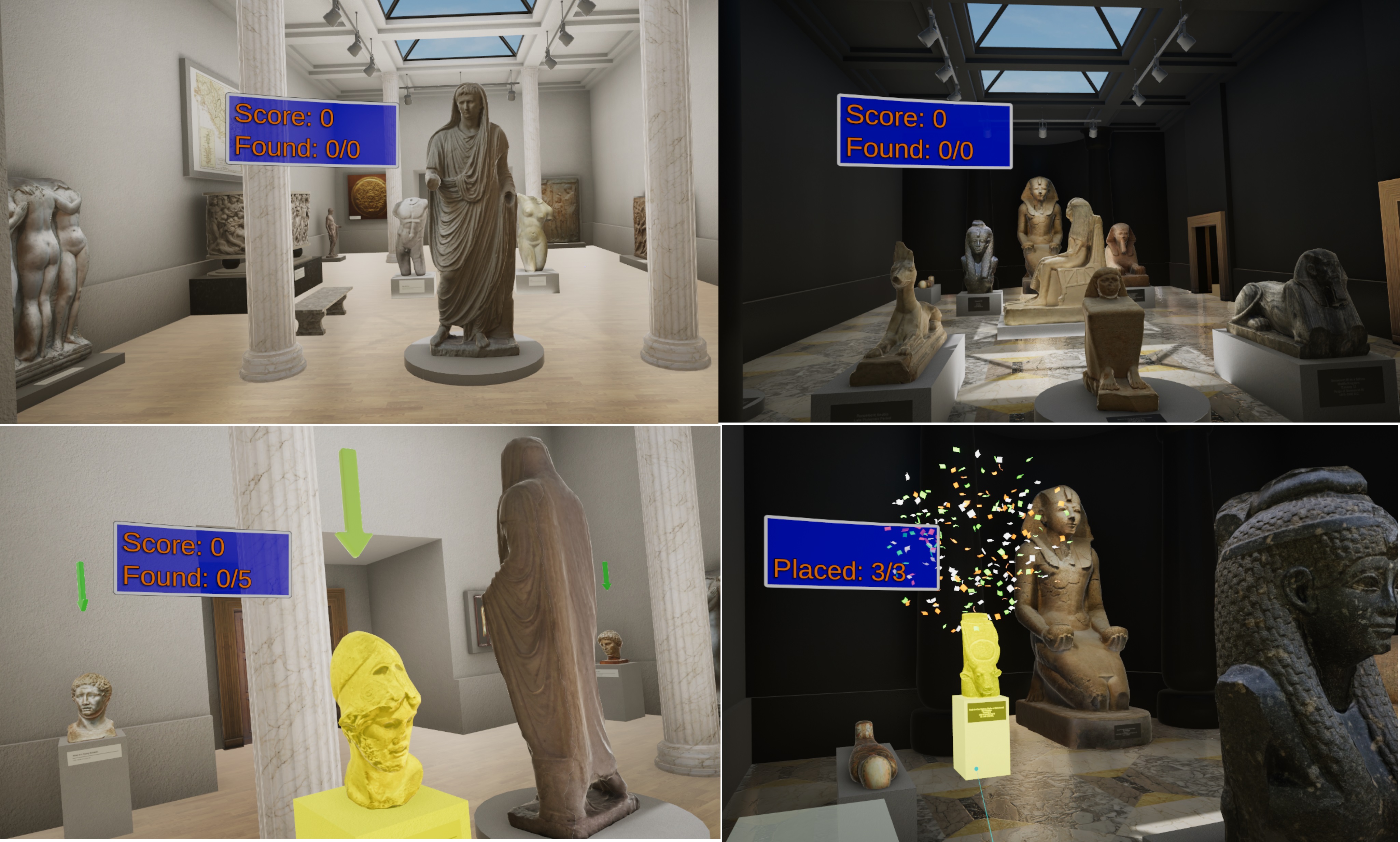}\\
	\end{tabular}
	\caption{The custom, multi-room, virtual museum environment, in which EM tasks are carried out. More rooms are unlocked as the difficulty level increases. Green arrow markers indicating which displays are to be remembered during the encoding phase. Bottom right: visual feedback for correct answer during the retrieval phase.}
	\label{fig:museum}
	 \vspace{-4mm}
\end{figure}
%%%%%%%%%%%%%%%%%%%%%%%%%%%%%%%%%%%%%%%%%
%Fig. 9.	The custom, multi-room, virtual museum environment, in which EM tasks are carried out. More rooms are unlocked as the difficulty level increases.
The EM task takes place in a multi-room virtual museum environment (see Fig.~\ref{fig:museum}). The core task is divided into two consecutive phases: encoding (storage of information, such that it can be distinguished from other distinct pieces of information) and retrieval (recognition of previously stored information) \cite{wang2012a}. In the encoding phase, players are asked to search for one or more displays at randomly generated locations in the museum. 
%These displays are marked with green arrows (see Fig. 10) to give players a visual indicator of what they are looking for. 
Players interact with marked displays in this phase using a laser pointer. On doing so, the age of the display is shown to the player. This interaction can take place at a distance, allowing a greater degree of spatial context to be encoded. After all the marked displays have been interacted with, the game transitions to the retrieval phase removing the marked displays from the museum, and teleporting the player back to the museum entrance.
%
%
%Fig. 10.	Markers and visual feedback. Top: green arrow markers indicating which displays are to be remembered during the encoding phase. Left: visual feedback for correct answer during the retrieval phase. Right: visual feedback for incorrect answer during the retrieval phase.

In the retrieval phase, players are tasked with placing a subset of the displays they interacted with during the encoding phase back in their original positions. The player uses the laser pointer to indicate where in the environment (from a selection of highlighted zones) they think it was located. 
%Correct placements award 5 points to the player's score, whereas incorrect placements deduct 4 points. 
On completion of the retrieval phase, a short bonus phase is initiated. Players are shown three displays they have interacted with and are asked which of them is the oldest/youngest. 
%Correct answers award 10 points. 
This textual (age) recall is not randomised, and players who can efficiently store the information in their LTM should perform better over repeated sessions.

\subsection{Data Acquisition}
After the study procedure and protocol was approved by Cambridge's Department of Computer Science and Technology Ethics Committee, 18 participants (5 female and 13 male, ranging in age from 20 to 37) volunteered to engage in the EMG data acquisition study, by wearing the HTC Vive VR headset with Faceteq sensing HMD foam insert (see Fig.~\label{fig:faceteq}). 
6 of these participants engaged in a preliminary pilot study, while the data collected from the remaining 12 formed the final annotated facial EMG dataset. 
%The participants were students and researchers recruited from the Cambridge Computer Science Department. 
%
%
%%%%%%%%%%%%%%%%%%%%%%%%%
\begin{figure}[!t]
	\begin{tabular}{c}
		\includegraphics[scale=0.55]{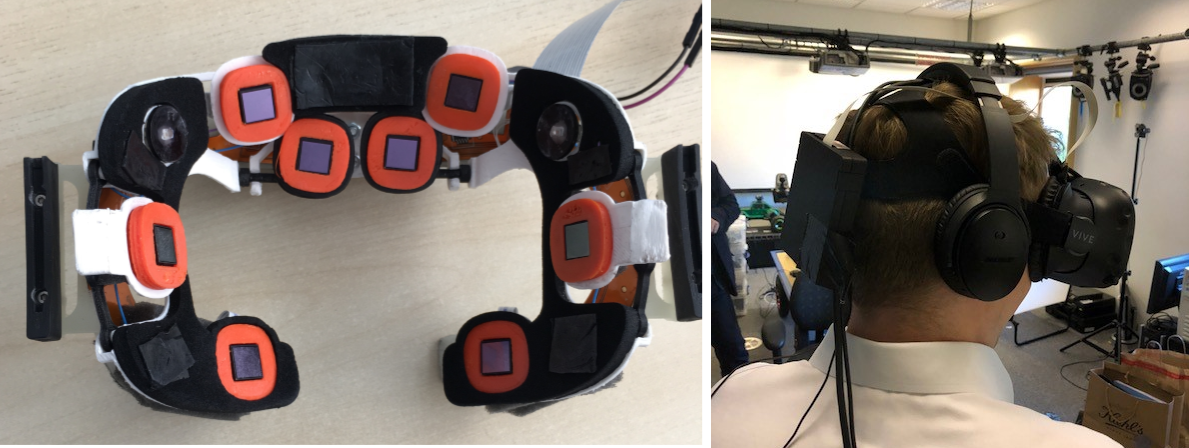}\\
	\end{tabular}
	\caption{Prototype Faceteq sensing HMD foam insert and how it is placed in the HTC Vive.}
	\label{fig:faceteq}
	 \vspace{-4mm}
\end{figure}

The participants were first acquainted with the research goals of the study through an information sheet and verbal introduction. 
They were introduced to the meaning of arousal and valence, and shown the affective slider annotation scheme (see Fig.~\ref{fig:slider}). The EMG recording sessions lasted for 3 minutes and 45 seconds, of which there were six in total (one per difficulty level, for both the WM and EM tasks). Half of the participants played through the difficulty levels in reverse order (hard-to-easy) to reduce the likelihood of the collected data being skewed positive (as in \cite{shumailov2017a}). The hypothesis here is based on the concept of the difficulty curve, the idea that, for an optimal experience, a game's difficulty should progress in a manner consistent with real-world skill acquisition (easy challenges during the cognitive stage,  moderate challenges during the associative stage, and more difficult challenges during the autonomous stage) \cite{b2009a}. By delivering challenges to the player in a reversed order, it is expected that they will experience negative affect more frequently (e.g. frustration early on, and boredom towards the end).

Participants were given a two minute break between gameplay sessions, allowing them to return to a neutral affective state. During these breaks, participants were asked to give an affective label that best summed up that session (using Russell's circumplex model \cite{russell1980a}). 
%This was done to check for any anomalies with the labels acquired in-game and to investigate the differences between mid and post-session annotation schemes. 
A short informal interview was conducted, after the EMG data collection and all other results from the study were recorded,  with the following questions. 
%This was to garner qualitative feedback that might inform future developments. The following questions were asked:

\begin{itemize}
\item Which of the two environments did you prefer spending time in?
\item Which of the two tasks did you find more engaging?
\item Did you experience any discomfort during the session and, if you have prior experience of VR, was the addition of the Faceteq sensor off-putting in any way?
\item To what extent, if any, did the annotation scheme affect your gameplay experience?
\end{itemize}
%%%%%%%%%%%%%%%%%%%%%%%%%
%\begin{figure}[!t]
%	\begin{tabular}{c}
%		\includegraphics[scale=0.3]{EMG-sensors.png}\\
%	\end{tabular}
%	\caption{EMG sensor locations top-to-bottom: Corrugator supercilii (right/left); Brow (right/left); Eye (right/left); Smile (right/left).}
%	\label{fig:sensorloc}
%\end{figure}
%%%%%%%%%%%%%%%%%%%%%%%%%%%%%%%%%%%%%%%%%

\subsection{Findings}
The majority of participants (10/12) expressed a preference for the museum environment over the supermarket, with many responses indicating that the supermarket felt more mundane as it is an environment they are overly familiar with in the real world. This may point to a trade-off between ecological validity and engagement in the choice of cognitive training environment. Responses were evenly split when it came to task preference, with many stating they preferred the EM task as it had more gameplay variety, while others appreciated the more naturalistic interactions in the WM task. %(e.g. physically touching items to pick them up, leaning across shelves/stalls, etc.). 
The response to the Faceteq sensor was positive. None of the participants indicated that they experienced any motion sickness or that the sensor was off-putting. Of the nine participants with prior VR experience, five particpants responded that they weren't aware of the sensor once they started playing, two participants responded that they there were aware of the sensor but it had no significant impact on their engagement, and two participants (self-identified regular VR users) stated that the ADC box attached to the back of the HMD (see Fig.~\ref{fig:faceteq}) served as a counterweight to the front-heavy HTC Vive. Most participants (7/12) stated that the in-game annotation scheme was mildly disruptive to the gameplay experience, while others either found it did not affect their experience (3/12) or found it very disruptive (2/12).

\section{SYSTEM EVALUATION}
This evaluation aims to investigate whether and how the developed CT environment and the overall system (with its modules for sensing, feature extraction and affect recognition) work together. 

\subsection{EMG Feature Extraction}
\subsubsection{Pre-processing} Sousa and Tavares noted, in their review of EMG normalisation methods \cite{sousa2012a}, that the voltage potential of surface EMG depends on several factors, varying between individuals and also over time within an individual. 
Baseline normalisation (removal) is a viable strategy to respond to these issues, having seen use in numerous studies on a variety of physiological signals (e.g., \cite{m2013a}). The user's baseline was read during the first 45 seconds of gameplay as at this point the novelty of VR had diminished to some extent, and the activities during the early-game are typically less arousing (e.g. reading shopping list, and looking at displays). 
%The second step is application of the normalization. 
%
%For the purposes of this study, a simple baseline division was employed.
\subsubsection{Feature Extraction} 
\begin{figure}[!t]
	\begin{tabular}{c}
		\includegraphics[scale=0.5]{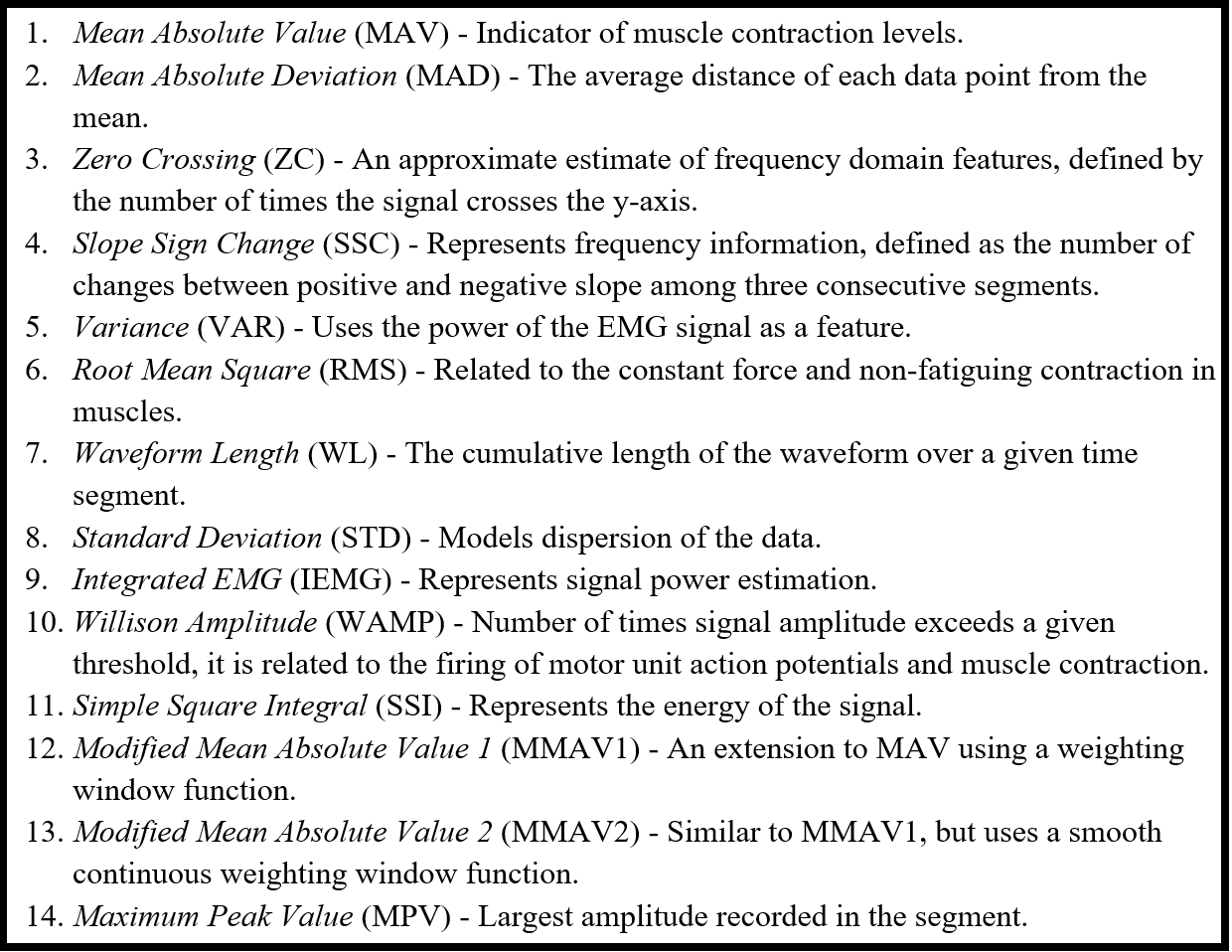}\\
	\end{tabular}
	\caption{List of time and time-frequency domain features extracted from the DWT approximation of the signal.}
	\label{fig:features}
	 \vspace{-4mm}
\end{figure}

After baseline normalisation, the input signal was processed using DWT \cite{zhang2010a}. The choice of mother wavelet for signal approximation was informed by the work of Phinyomark et al. \cite{phinyomark2010a}. They found that, for the purpose of denoising, coif5, Haar (db1), bior1.1 and rbio1.1 are the most suitable. A preliminary evaluation with our dataset showed that each of these wavelets resulted in very similar classification improvements (around +5\% to +7\% accuracy depending on the classifier). Therefore, going forward, the presented results are based on the DWT-Haar approximation of the EMG signal due to its efficient computation. 
A significant number of studies, in the domain of facial EMG classification, have shown temporal features to be the most informative \cite{jerritta2014a, perusqu2017a, hamedi2018a, soon2017a}. 
Based on these findings, the time and time-frequency domain features were extracted from the DWT approximation of the signal. These are shown in Fig. \ref{fig:features} (see \cite{tkach2010a} for mathematical definitions). Extracting this many features, from eight EMG channels, results in a high dimensional (8 * 14 = 112) dataset.  

\subsubsection{Feature Selection} 
Due to the dimensionality of the dataset, it was considered pertinent to include a feature selection step prior to classification. The strategy employed here was inspired by the work of 
Clerico et al. \cite{clerico2016a}, who utilised the minimal-redundancy-maximal-relevance criterion (mRMR) \cite{peng2005a} to select the best features in an EMG affective gaming context. 
mRMR attempts to find optimal features, based on mutual information, through forward selection.
The greatest classification improvement was achieved by selecting the best 30 features, identified by mRMR (around +3\% to +5\% accuracy depending on the classifier). These features were distributed among different muscle groups with the top 30 ranking made up of 10 features from users' eyes,  9 from their mouth, 7 from their eyebrows, and 3 from their corrugator supercilii.
%%%%%%%%%%%%%%%%%%%%%%%%%

\subsubsection{Findings}
SSC was the most common feature in the ranking, being extracted from all muscle groups bar the corrugator supercilii. SSC, extracted from the right eye sensor, was also computed to be the second most informative feature in the ranking. This was accompanied by MMAV1 extracted from the right mouth sensor (1st) and ZC extracted from the left mouth sensor (3rd), in a top three that scored significantly higher (by a factor of at least 3) than the remaining 27 features. Interestingly, extracted RMS features were considered relatively uninformative despite it being regularly cited as one in facial EMG emotion recognition research (e.g. \cite{jerritta2014a, hamedi2018a}). Though deviations in expected results, such as this, may be attributed to the fact that 
%this is the first study to utilise facial EMG in a VR context, and 
physical expressions of affect in peoples' faces are likely to be impacted by the VR headset.
%%%%%%%%%%%%%%%%%%%%%%%%%
\begin{figure}[!t]
\begin{tabular}{c}
	\includegraphics[scale=1.1]{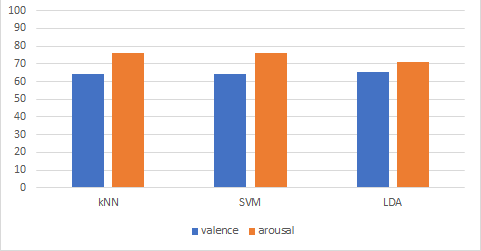}\\
	\end{tabular}
    \caption{Classification accuracies for kNN, SVM and LDA for positive valence vs. negative valence and high arousal vs. low arousal. Results are obtained using leave-one-subject-out cross-validation strategy.
}
	\label{fig:classification}
	 \vspace{-4mm}
\end{figure}
%%%%%%%%%%%%%%%%%%%%%%%%%%%%%%%%%%%%%%%%%
%\vspace{-1mm}
\subsection{Affect Classification}
%The distribution of (self-annotated) labels, acquired through the study, can be seen in Fig. 14 below. 
\subsubsection{Classification}
First the original (-1 to +1 continuous) valence/arousal labels were truncated into one of four emotion labels: energetic-positive (high valence, high arousal); calm-positive (high valence, low arousal); energetic-negative (low valence, high arousal); calm-negative (low valence, low arousal). This allows emotion recognition to be framed as a 4-class classification problem.
%Fig. 14.	Bar chart of acquired labels for valence and arousal (truncated for the 4-class classification problem). Y-axis represents the frequency of each label.
%
%
Three classifiers, which have been utilised to varying degrees of success in existing facial EMG literature \cite{jerritta2014a, perusqu2017a, hamedi2018a, soon2017a}, SVMs (with Gaussian kernel); 
kNN (with various values for k); and LDA were evaluated using the (subject-independent) leave-one-subject-out (LOSO) cross-validation strategy. The best classification results for each classifier can be seen in Fig.~\ref{fig:classification}.
The kNN (k = 4) offered the best arousal classification rate (76.2\%), and the best classification rate on the combined valence/arousal 4-class classification problem (68.8\%). LDA yielded the best valence classification rate (64.1\%), but fell behind kNN on the 4-class classification problem (65.9\%). The Gaussian SVM generally underperformed, though by investigating different kernels and further tuning hyperparameters, this could be improved in future work. 
%
%\begin{table}[htbp]
%\caption{Classification Accuracy shown for Valence / Arousal (\%)}
%\begin{center}
%\begin{tabular}{|c|c|c|c|}
%\hline
%\textbf{Classifier}&\multicolumn{3}{|c|}{\textbf{SVM kNN LDA}} \\
%\cline{2-4} 
%\textbf{Head} & \textbf{\textit{Table column subhead}}& \textbf{\textit{Subhead}}& \textbf{\textit{Subhead}} \\
%\hline
%64.1 / 76.2 & 64.1 / 76.2 & 65.6 / 71.1 \\
%\hline
%\multicolumn{4}{l}{$^{\mathrm{a}}$Sample of a Table footnote.}
%\end{tabular}
%\label{tab1}
%\end{center}
%\end{table}
\subsubsection{Findings}
The most noticeable discrepancy between the labels acquired in-game and those acquired post-game, was a consistently lower annotated value for arousal in the post-game interview. This manifested as a significantly lower classification accuracy for arousal when using the post-game labels as the four classes (around -15\% to -19\% depending on the classifier), while valence classification remained comparable. This suggests that the primary benefit of acquiring affective labels in-game is a more reliable estimate of the intensity of emotions. Placing participants in a reversed difficulty group had the desired effect of inducing negative affect more frequently (resulting in a more balanced dataset), with that group accounting for approximately 61\% of energetic-negative and 67\% of calm-negative annotations.
A notable difference between how our classification results were arrived at compared to those discussed in similar works (section 3.3), is that these results were computed using the (subject-independent) leave-one-subject-out (LOSO) cross-validation strategy. 
%This methodology has been shown to avoid non-independency issues that lead to unrealistic estimates of the generalisability of the model \cite{esterman2010a}. 
%%%%%%%%%%%%%%%%%%%%%%%%%
%\begin{figure*}
%\begin{figure}[!t]
%	\begin{tabular}{c}
%		\includegraphics[scale=1.70]{responses.jpg}\\
%	\end{tabular}
%	\caption{Post-game GEQ \cite{ijsselsteijn2007a} module responses for each participant (P). Values are on a scale of 0 (not at all) to 4 (extremely), and were calculated by averaging across their respective components.}
%	\label{fig:responses}
%\end{figure}
%\end{figure*}
%%%%%%%%%%%%%%%%%%%%%%%%%%%%%%%%%%%%%%%%%
%\begin{table}
%\begin{landscape}
%\begin{tabularx}{290mm}{|c|c|c|X|}
\begin{table*}[]
    \centering
     \setlength\tabcolsep{3.9pt}
    % \hspace*{-4mm}
    \fontsize{10}{11.5}\selectfont
    % \small
 %   \begin{tabular}{|C|L|L|}
 
%\renewcommand{\arraystretch}{1.2}
   \begin{tabularx}{\linewidth}{c|c|c|c|c|c|c|c|c|c|c|c|c|c|c}
 \hline
  %\multicolumn{2}{c|}
    \rowcolor{blue!10}
  {\textbf{PID}} &\multicolumn{2}{c|}{\textbf{P1}} & \multicolumn{2}{c|}{\textbf{P2}} &\multicolumn{2}{c|}{\textbf{P3}} &\multicolumn{2}{c|}{\textbf{P4}} &\multicolumn{2}{c|}{\textbf{P5}} &\multicolumn{2}{c|}{\textbf{P6}} &\multicolumn{2}{c}{\textbf{Mean}} \\
\hline
  \rowcolor{blue!10}
  
 \textbf{Game Version} & \textbf{Ad} & \textbf{NAd} & \textbf{Ad} & \textbf{NAd} & \textbf{Ad} & \textbf{NAd} & \textbf{Ad} & \textbf{NAd} & \textbf{Ad} & \textbf{NAd} & \textbf{Ad} & \textbf{NAd} & \textbf{Ad} & \textbf{NAd} \\
 Competence	&	2	&	1	&	3	&	2.5	&	2.5	&	0.5	&	3	&	2	&	2	&	2	&	3	&	2	&	2.58	&	1.67	\\ \hline
Sensory and Imaginative Immersion	&	4	&	3	&	4	&	4	&	3	&	2	&	3	&	3	&	2	&	3	&	4	&	4	&	3.33	&	3.17	\\ \hline
Flow	&	3.5	&	3.5	&	4	&	3.5	&	3	&	2	&	3	&	3	&	2	&	3	&	3.5	&	2	&	3.17	&	2.83	\\ \hline
Tension	&	2	&	3.5	&	0.5	&	0.5	&	0	&	2	&	0	&	0.5	&	2	&	0	&	0.5	&	1.5	&	0.83	&	1.33	\\ \hline
Challenge	&	4	&	4	&	2.5	&	3	&	3	&	4	&	3.5	&	4	&	1	&	3	&	2	&	3.5	&	2.67	&	3.58	\\ \hline
Negative Affect	&	1.5	&	3	&	0	&	0.5	&	0	&	1	&	0	&	0	&	0.5	&	0	&	0	&	0	&	0.33	&	0.75	\\ \hline
Positive Affect	&	2.5	&	1.5	&	3.5	&	3.5	&	3	&	2	&	2.5	&	2	&	1.5	&	3	&	3	&	2.5	&	2.67	&	2.42	\\ \hline
\end{tabularx}
%\end{landscape}
     \caption{In-game GEQ \cite{ijsselsteijn2007a} module responses for each participant (P) represented with participant ID (PID), for Adaptive (Ad) and Non-Adaptive (NAd) versions of the game. Values are on a scale of 0 (not at all) to 4 (extremely), and were calculated by averaging across their respective components.}
      \label{tab:IG-evaluation}
    \vspace{-4mm}
\end{table*}
%%%%%%%%%%%%%%%%%%%%%%%%%%%%%%%%%%%%%%%%%
%%%%%%%%%%%%%%%%%%%%%%%%%%%%%%%%%%%%%%%%%
\begin{table}[h!]
\centering
\begin{tabular}{ c|c|c|c|c|c|c|c } 
\hline
  \rowcolor{blue!10}
\textbf{PID}	& \textbf{P1} &	\textbf{P2} & \textbf{P3}	 & \textbf{P4}  & \textbf{P5}	 & \textbf{P6} & \textbf{Mean}\\
\hline
Positive	& 2.4 &	2.6	& 1.4 &	1.6 & 2 &	2.8	& 2.13 \\
\hline
Negative	& 0.33 &	0 &	0 &	0 &	0 &	0.33 &	0.11\\
\hline
Tired &	1 &	0.5	 & 0 &	2.5 &	0 &	0 &	0.67\\
\hline
R2R &	1 &	1 &	0.33 &
0.33 &	0.33 &	1.66 &	0.78\\
\hline
\end{tabular}
\caption{Post-game GEQ module responses for each participant (P) on a scale of 0 (not at all) to 4 (extremely), calculated by averaging across their respective components. R2R refers to Return to Reality.}
      \label{tab:PG-evaluation}
 \vspace{-4mm}
\end{table}
%%%%%%%%%%%%%%%%%%%%%%%%%%%%%%%%%%%%%%%%%
%
%\vspace{-1mm}
\section{Intervention Evaluation}
The goal of this evaluation was to gather qualitative feedback from the target population (older adults) to examine the potential benefits of utilising affective adaptation for CT and gather insights for defining future research steps. 

\subsection{Affective Feedback Loop Integration}
%The motivation for utilising DDA in video games was stated succinctly by Hunicke, games are boring when they are too easy, but frustrating when they are too difficult \cite{hunicke2005a}. 
The system integrated here relies on both affect sensing and player performance as a data point, due to its exhibited value in existing DDA solutions and to offset the (classification) inaccuracies in the model (i.e. to avoid fixing what is not broken \cite{hunicke2005a}).
The number of difficulty levels in this new adaptive version of the game was increased from three (easy-medium-hard) to ten (10-point scale). This allows for more subtle transitions in difficulty to avoid the player becoming overly conscious of the adaptation (and potentially feeling `cheated' by it \cite{hunicke2005a}). Every 45 seconds (in place of the annotation interface from the previous study) the player's affect is classified in real-time based on incoming facial EMG signals. The following DDA rules, encompassing both player affect and performance, govern how the difficulty is adapted:
%\begin{itemize}
%\item \textit{[Calm-negative + perfect score]:} increment difficulty by 2; 
%\item \textit{[Calm-negative + imperfect score] or [positive valence + perfect score]:} increment difficulty by 1;
%\item \textit{[Positive valence + imperfect score]:} no change in difficulty;
%\item \textit{[Negative score] or [energetic-negative + imperfect score]:} decrement difficulty by 1;
%\item \textit{[Energetic-negative + negative score]:} decrement difficulty by 2.
%\end{itemize}
\\
\\
\textit{[Calm-Negative + :Perfect Score]:} Increment difficulty by 2;\\ 
\textit{[Calm-Negative + Imperfect Score] or [Positive Valence + Perfect Score]:} Increment difficulty by 1;\\
\textit{[Positive Valence + Imperfect Score]:} No change in difficulty;\\
\textit{[Negative Score] or [Energetic-Negative + Imperfect Score]:} Decrement difficulty by 1;\\
\textit{[Energetic-Negative + Negative Score]:} Decrement difficulty by 2.\\
%\vspace{-1mm}

This rule set was arrived at following a short testing period with pilot participants. For instance, in the  
\emph{[Calm-Negative + Perfect Score]} case, it is hypothesised that the player is not enjoying the game as valence is negative, and is likely bored as they are in a non-energetic state, and the score is perfect, so the level is increased by two. Instead, \emph{[[Positive Valence + Perfect Score]} leads to a smaller increment of one in order to gradually maximise the cognitive challenge provided by the game, while minimising the risk of disrupting the player’s engagement (as indicated by the positive valence score). In the best-case execution, the adaptation is intended to lead players to a flow state \cite{csikszentmihalyi2014a}, where they are faced with tasks that they have a chance of completing through application of their skills. In addition to being a signifier of high engagement, being in a state of flow has been shown to improve cognitive performance \cite{gabana2017a}.

%\vspace{-4mm}
\subsection{Evaluation Study}
The protocol for the evaluation study was largely similar to the previous study, with a few notable exceptions. 6 participants (4 female and 2 male, ranging in age from 60 to 100), with no history of cognitive impairment, volunteered to engage in this evaluation study. The recruitment of this older control group was facilitated by the researchers from The University of Cambridge's Department of Neuroscience. None of the participants played video games with any degree of regularity and only one had prior experience of VR. 
%Department of Clinical Neurosciences in Cambridge. 
%Participants were first acquainted with the research goals of the study through an information sheet and verbal introduction. 
%They were then informed of their rights, and signed consent was obtained. 
Participants started by completing a battery of standardised cognitive tests 
\cite{mioshi2006a, nelson1982a, osterrieth1944a, crockett2008a, tombaugh2004a, chan2016a, wechsler1958a} (administered by the Clinical Neuroscience researchers). This enabled accurate characterisation of the sample and the investigation of correlations between the standardised tests and the new VR paradigm. The time taken to administer these tests averaged at about 1 hour.

Participants were given an explanation of the task goals and time to practice in the VR environment prior to starting the session proper. They played both an adaptive and non-adaptive (linearly increasing difficulty) version of the game (without being told which version is which), in two, fifteen minute, gameplay sessions (7 minutes and 30 seconds for both WM and EM tasks). To mitigate any order effects bias in the evaluation of the adaptive and non-adaptive versions, half of the participants played the adaptive version first, while the other half played the non-adaptive version first. Participants' subjective experience (i.e. immersion, engagement, and flow) with the VR training intervention was evaluated using the in-game and post-game components of the game experience questionnaire (GEQ) 
\cite{ijsselsteijn2007a}. 

Each session was concluded with an informal interview that took place after all other results from the study were recorded. The contents of this interview was the same as in the first study, with the exception of an additional question on the impact of the in-game affective annotation during the first study. 
%The purpose of the interview was to garner further, open-ended, qualitative feedback (from the target user group of older adults), and further characterise the sample with knowledge of their prior experience of VR and video games. 
%
Participation in the study lasted for about 2 hours and 15 minutes on average (including breaks).
Immediately after playing each version (adaptive/non-adaptive) of the game, participants reported on their feelings of competence, sensory and imaginative immersion, flow, tension, challenge, negative affect, and positive affect by completing the \emph{in-game module} of the GEQ. Each of these categories is represented by a series of sub-components in the questionnaire, a numeric value is then computed for each by averaging across their sub-component values. The accumulated responses  can be found in Table~\ref{tab:IG-evaluation}. Finally, when both gameplay sessions were complete, the participants filled out the \emph{post-game module} of the GEQ \cite{ijsselsteijn2007a}. This gave participants the opportunity to think and reflect on the experience as whole. The accumulated results, similarly calculated by averaging across their sub-components in the questionnaire, are provided in Table~\ref{tab:PG-evaluation}.

\subsection{Findings}
Looking at Table~\ref{tab:IG-evaluation}, while the response to both versions of the game can generally be described as positive, there are a few noteworthy differences.
The two standout differences are the increased feeling of competence and the decreased feeling of challenge while playing the adaptive version of the game. The noteworthy increase in competence is particularly encouraging as it relates to one of the key deficiencies identified with existing cognitive training interventions, the drop-off in user engagement 
\cite{Simons2016}. In their research of intrinsic motivation, Deci and Ryan argue that structures that enable feelings of competence during action can enhance intrinsic motivation for that action \cite{deci1985a}. This increased feeling of competence, brought about through affective adaptation, highlights the potential of adaptive techniques in motivating users to engage with cognitive training interventions. However, specific aspects related to intrinsic motivation were not included in the questionnaires employed in our studies. For more insightful conclusions on intrinsic motivation, relevant aspects should be investigated explicitly in future studies.

While the drop in challenge is not unequivocally positive, the decrease to a more neutral value in the adaptive version, along with the slight increase in flow (which describes a state of high engagement), suggests that participants are being met with more appropriate challenges that they can overcome using their skills \cite{csikszentmihalyi2014a} (potentially explaining the slight increase in positive affect, and decrease in negative affect while playing the adaptive version).
The largely positive responses provided for the \emph{post-game module} of the GEQ as seen in Table~\ref{tab:PG-evaluation}, in conjunction with those recorded by the \emph{in-game module}, are promising indicators that gamification and VR can play a role in increasing engagement with cognitive interventions in older adults.

Another area of interest for this evaluation was to what extent the WM and EM tasks, implemented in VR, engaged the intended cognitive abilities of the participants. This was examined by looking for correlations between how participants performed (relative to each other) in the standardised tests and the VR tasks. Positive correlations, calculated using Spearman's rank-order correlation (rho), were found between the performance rankings of participants in the VR paradigm and closely related standardized tests. Most notably, high positive correlations were found between the Trail Making Test \cite{tombaugh2004a}, which examines executive functioning (a superset of WM), and the WM task in VR (rho=+0.60), and between the 4 Mountains test (a short SM test) and the EM task in VR (rho=+0.74).

In the post-session interview, all 6 participants stated their preference for the museum environment (finding the supermarket more mundane). The overall consensus was, however, that the environments they would prefer would match those in the real-world. 5 participants preferred the EM task, stating that while they found it to be more complex, the greater variety it offered was a motivating factor for them to return to it and improve. This may point to task variety being an important factor in maintaining user engagement in cognitive training. All 6 participants responded that they felt no motion sickness during the study. 4 out of the 6 participants indicated that they felt no facial discomfort, while 2 participants, who wore glasses throughout the study, felt a bit of pressure on their face towards the end. This was likely a result of the slightly thicker face cushion used with the Faceteq prototype.
%(compared to the default HTC Vive cushion). 
%
%All participants responded that the weight of HMD, and attached ADC box, did not bother them (1 participant added that it took time to get used to it, and another that they would like regular breaks if they were to use it for longer durations). 
%
%The participants were then encouraged to give more open-ended feedback. The blurriness of the low resolution VR lenses was reported by 2 participants as having a negative impact on their performance and immersion. 
%This is a deficiency shared by most current generation VR devices and will only be solved with future hardware releases. 
%
\section{Summary and Conclusion}
This work investigated 
%the application of a variety techniques, and technologies, in 
the development of
%more engaging cognitive training schemes. It has detailed the steps in the process for production of 
an affect-aware VR game for cognitive training using facial EMG signals for affect classification.
%
%Across 18 participants (ranging in age from 20 to 100), our choice of sensor (Faceteq prototype), was universally considered to be unobtrusive (a much cited flaw in facial EMG applications).
%Our experiments demonstrated that this physiological modality can be relied on, with a relatively small data source (12 participants, with 29 labeled 45 second EMG segments per participant), to moderate success. 
Classification rates of 64.1\% and 76.2\%, for valence and arousal respectively, were achieved through a combination of %baseline normalisation, 
DWT-Haar filtering, temporal feature extraction, feature selection, and kNN classification.
%The combined valence/arousal 4-class classification problem served as a reasonably accurate driver (in unison with performance-driven adaptation) for affective adaptation, with an emotion recognition rate of 68.8\% (as determined by subject independent LOSO CV).
%
The promise of DDA in the development of more engaging cognitive training was substantiated through a small-scale user study with older adults. 

The qualitative feedback garnered over the course of the study pointed to a notable increase in feelings of competency 
%(associated with intrinsic motivation and engagement \cite{deci1985a}), 
and participants being more appropriately challenged. %(moderate feelings of challenge, and a slight increase in flow). 
%Strong positive correlations were found between the participant rankings obtained in the baseline neuropsychology tests and those obtained in the VR paradigm, suggesting that performance in the VR training tasks may be a good indicator of cognitive capacity. 
Participant feedback relating to both the adaptive and non-adaptive versions of the game was largely positive. This response lends credence to the notion that gamification and VR are viable tools for improving engagement in cognitive training with older adults. The findings here should be qualified by reiterating an inherent limitation of the study. Six participants is a small user group for an evaluation study 
%(necessitated by the constrained timescale for the work) 
and should be expanded in future research to fully determine the veracity of these findings.\\
\\
%
%In short, this work represents a preliminary step in fully ascertaining the benefits of gamification, virtual reality and dynamic adaptation techniques for cognitive training interventions, and there is still much work to be done.
%While the response and feedback to this study have been encouraging, it is important to qualify these results by reiterating the small sample size being used, and the once off nature of the study. Nevertheless, these are promising indicators which suggest that this is a domain that warrants further research.
%
%
%\begin{itemize}
%\item Use either SI (MKS) or CGS as primary units. (SI units are encouraged.) English units may be used as secondary units (in parentheses). An exception would be the use of English units as identifiers in trade, such as ``3.5-inch disk drive''.
%\item Avoid combining SI and CGS units, such as current in amperes and magnetic field in oersteds. This often leads to confusion because equations do not balance dimensionally. If you must use mixed units, clearly state the units for each quantity that you use in an equation.
%\item Do not mix complete spellings and abbreviations of units: ``Wb/m\textsuperscript{2}'' or ``webers per square meter'', not ``webers/m\textsuperscript{2}''. Spell out units when they appear in text: ``. . . a few henries'', not ``. . . a few H''.
%\item Use a zero before decimal points: ``0.25'', not ``.25''. Use ``cm\textsuperscript{3}'', not ``cc''.)
%\end{itemize}

%\\
%\textbf
 %\vspace{-4mm}
%\subsection{ACKNOWLEDGEMENTS}
\textbf{ACKNOWLEDGEMENTS.} This work has been partially supported by the EPSRC (grant ref. EP/R030782/1).

%\newpage
%\balance

%\bibliographystyle{ACM-Reference-Format}
%\bibliography{References.bib}

%% End of file `sample-sigconf.tex'.

%\newpage
\balance

\bibliographystyle{IEEEtran}
\bibliography{References.bib}
% \nocite{*}
\end{document}